\documentclass[twocolumn,aps,preprintnumbers,amsmath,amssymb,superscriptaddress,nofootinbib]{revtex4}
\usepackage{graphicx}
\usepackage{subfigure}
\usepackage{amsmath, amssymb}
\usepackage{multirow} 
\usepackage{hhline}

\newcommand{\lsim}{\lesssim}

\newcommand{\gsim}{\gtrsim}

\newcommand{\beq}{\begin{equation}}
\newcommand{\eeq}{\end{equation}}
\newcommand{\bea}{\begin{eqnarray}}
\newcommand{\eea}{\end{eqnarray}}

\newcommand{\eps}{\varepsilon}
\newcommand{\pb}{{\rm pb}}
\newcommand{\fb}{{\rm fb}}
\newcommand{\br}{{\rm BR}}
\newcommand{\tev}{{\rm TeV}}
\newcommand{\gev}{{\rm GeV}}

\newcommand{\met}{{/\!\!\!\! E_T}}

\begin{document}
\title{Dark decay of the top quark}
\author{Kyoungchul Kong}
\affiliation{Department of Physics and Astronomy, University of Kansas, Lawrence, Kansas 66045, USA}
\author{Hye-Sung Lee}
\affiliation{Department of Physics, College of William and Mary, Williamsburg, Virginia 23187, USA}
\affiliation{Theory Center, Jefferson Lab, Newport News, Virginia 23606, USA}
\author{Myeonghun Park}
\affiliation{Kavli IPMU (WPI), The University of Tokyo, Kashiwa, 277-8583, Japan}

\date{January, 2014}

\preprint{IPMU14-0012}

\begin{abstract}
We suggest top quark decays as a venue to search for light dark force carriers.
The top quark is the heaviest particle in the standard model whose decays are relatively poorly measured, allowing sufficient room for exotic decay modes from new physics.
A very light (GeV scale) dark gauge boson ($Z'$) is a recently highlighted hypothetical particle that can address some astrophysical anomalies as well as the $3.6 \sigma$ deviation in the muon $g-2$ measurement.
We present and study a possible scenario that top quark decays as $t \to b W + Z's$.
This is the same as the dominant top quark decay ($t \to b W$) accompanied by one or multiple dark force carriers.
The $Z'$ can be easily boosted, and it can decay into highly collimated leptons (lepton-jet) with large branching ratio.
We discuss the implications for the Large Hadron Collider experiments including the analysis based on the lepton-jets.
\end{abstract}
\maketitle

\section{Introduction}
With a recent discovery of a new scalar particle \cite{Aad:2012tfa,Chatrchyan:2012ufa}, which is consistent with the standard model (SM) Higgs boson, the understanding of the SM is near completion except for precision studies.
With this, the interest toward new physics beyond the SM soars.
One definite evidence of new physics is the existence of dark sectors such as dark energy and dark matter \cite{PDG}.

Dark force was introduced as a newcomer to dark sector \cite{ArkaniHamed:2008qn}.
It is a hypothetical interaction among the dark matters that can address various astrophysical anomalies such as positron excess.
Positron excess has been observed at numerous experiments including PAMELA \cite{Adriani:2008zr} and AMS \cite{Aguilar:2013qda}.
A very light (roughly, GeV scale) dark force carrier (which we call $Z'$) is supposed to couple to dark matter strongly, but extremely weakly to the SM particles.
It is expected to be light because such a light force carrier can provide the the required enhancement (via the so-called Sommerfeld effect) of the present time dark matter annihilation at the Galactic center while satisfying the dark matter relic density constraints \cite{ArkaniHamed:2008qn}.
Through a simple kinematics, it can also naturally explain why antiproton excess has not been observed.
In addition, such a gauge boson with small mass and very weak coupling can address the $3.6 \sigma$ deviation in the muon anomalous magnetic moment \cite{PDG} through a one-loop correction of $Z'$ \cite{Fayet:2007ua,Pospelov:2008zw}.
(For more motivations and details about the light dark force carriers, see Ref.~\cite{Essig:2013lka}.)

With such appealing motivations, there are active searches for light dark force carriers.
A dark force carrier is roughly of $\gev$ scale, and can be searched for both at the low energy and high energy experiments.
The major search schemes are based on bremsstrahlung at fixed target experiments or meson decays \cite{Bjorken:2009mm}.
Collider signatures for certain supersymmetric models \cite{ArkaniHamed:2008qp} and the decay of the Higgs bosons into the $Z'$ \cite{Davoudiasl:2012ag,Davoudiasl:2013aya,Lee:2013fda} have been studied as well.

In this paper, we present a novel channel to produce and search for the dark force carrier using top quark decays.
We show that the top quark decays into dark force carriers are possible and it can be searched for at the Large Hadron Collider (LHC) with excellent discovery potential.

\section{\boldmath Top to $Z'$ production}

\begin{figure}[t]
\begin{center}
\subfigure[]{
\includegraphics[height=0.17\textwidth,clip]{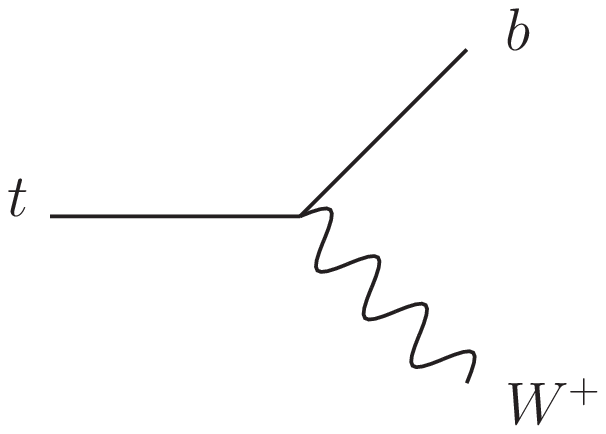}} ~
\subfigure[]{
\includegraphics[height=0.17\textwidth,clip]{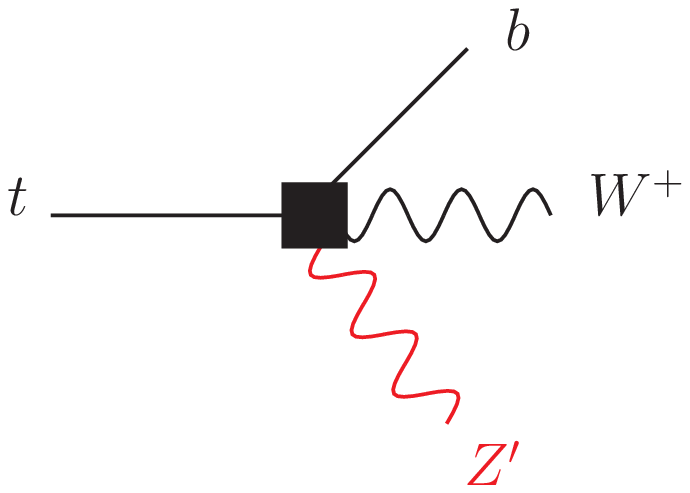}} 
\end{center}
\caption{(a) Dominant top quark decay mode ($t \to b W$). (b) The top quark decay into dark force carriers ($t \to b W + Z'$s) is the same to the dominant top quark plus one or multiple $Z'$s.
A very light $Z'$ can be easily boosted and, it can decay into highly collimated jets or leptons (lepton-jet).}
\label{fig:topdecay}
\end{figure}

The top quark is the heaviest elementary particle in the SM, quite possibly heavier than new particles in many models beyond the SM.
(For a short review about the top quark, see a review article in Ref.~\cite{PDG}.)
Because of its large mass, its lifetime ($\tau_t \approx 0.5 \times 10^{-24}$ sec) is very short.
Since it is shorter than the QCD scale ($\Lambda_\text{QCD} \sim 10^{-23}$ sec), top quark decays before forming any hadrons, unlike other quarks \cite{Bigi:1986jk}.
For this reason, top quark decays have been considered as an ideal probe of new physics even before its discovery at Tevatron in 1995 \cite{Abe:1995hr,Abachi:1995iq}.

Although top quark mass is precisely measured ($m_t = 173.1 ~\gev$ with $0.5 \%$ level uncertainty), its decay width is not \cite{PDG}:
\beq
\Gamma_t = 2.0 \pm 0.5 ~\gev .
\eeq
While it is in a reasonable agreement with the SM prediction ($\Gamma_t^\text{SM} \simeq 1.3 ~\gev$ with theoretical error better than $1 \%$ \cite{PDG}), quite a large uncertainty ($25\%$) indicates a lack of precise knowledge about top quark properties.
While the exact values are process dependent, typical experimental errors related to top quark decays are of ${\cal O} (10 \%$) \cite{PDG}, leaving plenty room for new decay modes originating from new physics beyond the SM.
(Later, in our illustration, we will take the top quark decay branching ratio into light $Z'$s not more than $0.1\% - 1 \%$ although it could be much greater in principle.)
With these observations, we view the top quark decay as an ideal window to look for a dark force carrier.

In the SM, nearly $100 \%$ of top quark decay is the on-shell $t \to b W$ decay.
Possible scenarios to search for $Z'$ in connection to the top quark decay include \\
\indent (i) $t \to b H^+ \to b W + Z'$ \\
\indent ~~~ (through $H^\pm W^\mp Z'$ coupling), \\
\indent (ii) $t \to b H^+ \to b W + h \to b W + Z' Z'$ \\
\indent ~~~ (with a light non-SM Higgs boson $h$), \\
\indent (iii) $t \to b W^* \to b W + Z'$ \\
\indent ~~~ (through $Z' W W$ coupling), \\
\indent (iv) $t \to b W^* \to b W + h \to b W + Z' Z'$ \\
\indent ~~~ (through $hWW$ coupling). \\
There can be also $Z'$ radiation off from top not being a decay product.
For a relatively heavy $Z'$, $m_{Z'} \gsim 1 ~\gev$, the radiation cross sections are negligibly small though.
While the off-shell processes are worth investigating as they may prevail in different situations (for example, when a charged Higgs is absent), we will focus on the on-shell decays in this paper.

There may be also other ways the $t$ can decay into $Z'$s without producing $b W$ such as $t \to q Z'$ (with $q = u, c$) through the $W$ loop.
Nevertheless, it is interesting to observe that there are abundant ways the $t$ can decay into $Z'$s with $b W$, the decay products of the dominant top quark decay. (See Fig.~\ref{fig:topdecay}.)

Throughout this paper, we will assume $m_W \lsim m_{H^\pm} \lsim m_t$, and study the on-shell decays (i) and (ii).
Whether (i) or (ii) dominates depends on the masses of the Higgs bosons, especially whether the non-SM Higgs $h$ is light enough so that the charged Higgs can decay into it dominantly or not.

\section{Ingredients of dark force models}
\subsection{Dark force carrier}
While the final particles in the aforementioned modes are the same up to the number of $Z'$s, some of the modes are model dependent, and it is worth describing some aspects of the models.

The minimum ingredients of extra particles to extend the SM to the dark sector are, besides the dark matter itself, the $Z'$ (dark force carrier) and an additional Higgs to give a mass to the $Z'$.
As the $Z'$ should be massive in order to decay into the leptons so that it can explain the astrophysical anomalies, we need some extended Higgs sector to give a mass to the $Z'$.
(We do not consider other possible ways to provide a mass such as the Stueckelberg mechanism \cite{Ruegg:2003ps}.)

$Z'$ is typically taken as a gauge boson of a new gauge symmetry, dark $U(1)$, under which the SM particles do not have charges.
Although $Z'$ does not couple to the SM particles directly, it can couple through the mixing of the $Z'$ with the SM gauge bosons via the gauge kinetic mixing parametrized by $\eps$ \cite{Holdom:1985ag}
\beq
{\cal L}_\text{gauge} = - \frac{1}{4} B_{\mu\nu} B^{\mu\nu} + \frac{1}{2} \frac{\eps}{\cos\theta_W} B_{\mu\nu} Z'^{\mu\nu} - \frac{1}{4} Z'_{\mu\nu} Z'^{\mu\nu} .
\eeq
The exact coupling, however, depends on the details of model, especially on how the $Z'$ gets a mass.
For example, it depends on whether the extra Higgs is a $SU(2)_L$ singlet or doublet \cite{Davoudiasl:2012ag}.

Depending on the Higgs sector, the $Z'$ may couple both to the electromagnetic current ($J_{em}$) and the weak neutral current ($J_\text{NC}$).
The interaction Lagrangian of the $Z'$ with the SM fermions is given by
\bea
{\cal L}_{\text{dark } Z} &=& - \left( \eps e J_{em}^\mu + \eps_Z g_Z J_\text{NC}^\mu \right) Z'_\mu \\
&=& \bar f \left( g_V \gamma^\mu - g_A \gamma^\mu \gamma^5 \right) f Z'_\mu
\eea
with
\bea
g_V &=& -\eps e Q_f - \eps_Z g_Z \left(\frac{1}{2} T_{3f} - Q_f \sin^2\theta_W\right) , \\
g_A &=& -\eps_Z g_Z \left( \frac{1}{2} T_{3f} \right) ,
\eea
where $Q_f$ and $T_{3f}$ are the electric charge and isospin, respectively.
Bounds on the couplings of the $Z'$ come from various experiments such as the lepton anomalous magnetic moment, atomic parity violation, polarized electron scattering, meson decays, fixed target experiments, beam dump experiments, and Higgs decays.
The exact bounds depend on the $m_{Z'}$ and its decay branching ratio, but typically, it is set as $|\eps| \lsim 10^{-2}$ and $|\delta| \lsim 10^{-2}$ (with $\eps_Z \equiv \delta \, \frac{m_{Z'}}{m_Z}$) \cite{Davoudiasl:2012ag,Davoudiasl:2012qa}.

For our interested $Z'$ of roughly ${\cal O}(\gev)$, the $\br(Z' \to \ell^+ \ell^-)$ for $\ell = e$, $\mu$ is expected to be large, typically $0.2 \sim 1$ depending on $m_{Z'}$ and other details.
Such a light $Z'$ gauge boson shows many distinguishable properties from heavy (electroweak or TeV scale) $Z'$ gauge bosons in various contexts \cite{Langacker:2008yv}.
(Also see Refs.~\cite{Alves:2013tqa,Arcadi:2013qia} for some recent studies on invisible or partly invisible heavy $Z'$ gauge bosons.)

Because the $Z'$ can be easily boosted in the top decay processes we consider, it is expected to appear as highly collimated leptons or jets.
Depending on its mass and coupling, it could be more identifiable as a lepton-jet or a simple pair of isolated leptons. (See the Appendix for some detailed discussions.)
A lepton-jet is a final state consisting of collimated electrons or muons.
Measuring properties of lepton-jets have been studied and experimental searches at the LHC experiments already started \cite{Aad:2012qua,Aad:2013yqp}.

\subsection{New Higgs bosons}
The dark force can be categorized into how the $Z'$ gets a mass.
We will consider the dark $Z$ model based on the type-I two Higgs doublet model (2HDM) \cite{Davoudiasl:2012ag}, with notations taken from Ref.~\cite{Lee:2013fda}, for scenarios (i) and (ii).
In this kind of dark force model, $\tan\beta \equiv v_2 / v_1 \gsim 1 $ is required, where $\Phi_2$ couples to the SM fermions and $\Phi_1$ does not.
(For a heavy $Z'$ model in 2HDM frames, for example, see Ref.~\cite{Ko:2013zsa}.)

A new neutral Higgs boson that can mix with the SM Higgs doublet is generically expected, and if an extra Higgs is a doublet, there are charged Higgs bosons too.
The Higgs properties of such scenarios have been studied in many literatures \cite{Davoudiasl:2012ag,Lee:2013fda,Chang:2013lfa}.
A new neutral Higgs can be very light and it can dominantly decay as $h \to Z' Z'$.
A charged Higgs can be also much lighter than typical experimental bounds as its dominant decay may be into rather elusive light $Z'$s \cite{Lee:2013fda}.

For the $m_{H^\pm} \lsim m_t$, the major decay modes of $H^+$ in the typical 2HDM are into $\nu \tau^+$ and $c \bar s$.
Their decay widths are
\bea
\Gamma (H^+ \to \nu \tau^+) &\simeq& \frac{m_{H^\pm}}{8 \pi v^2} \frac{m_\tau^2}{\tan^2\beta} ,
\eea
and similarly for $c \bar s$ with a color factor.
Despite the color factor, because of the small mass of charm quark ($c$) at the electroweak scale, the $c \bar s$ mode is subdominant to the $\tau \nu$ mode.
With a recently discovered $125 ~\gev$ SM-like Higgs ($H_\text{SM}$), the off-shell decay $H^\pm \to W^* H_\text{SM} \to f \bar f' H_\text{SM}$ can be also quite sizable, even larger than $H^+ \to \nu \tau^+$ for certain parameter space.

The scenario (i) is based on the assumption that the SM-like Higgs is the lighter Higgs doublet.
The $H^\pm W^\mp Z'$ coupling is very small, but its decay branching ratio could be sizable \cite{Ramos:2013wea,Davoudiasl:2014mqa}.
The tree-level decay width is given by
\beq
\Gamma (H^\pm \to W Z') \simeq \frac{m_{H^\pm}^3}{16 \pi v^2} \left( \sin\beta \cos\beta_d \right)^2 \left( 1 - \frac{m_W^2}{m_{H^\pm}^2} \right)^3
\eeq
where $\beta_d$ is a parameter related to the dark sector Higgs singlet, introduced in Ref.~\cite{Davoudiasl:2012ag}.

Scenario (ii) is based on the assumption that there is a light Higgs $h$ that a charged Higgs can decay into.
For definiteness, let us take the Higgs mixing angle $\alpha \simeq \pm \pi / 2$ limit, which is a decoupling limits of doublets where a heavier Higgs doublet is the SM-like Higgs ($125 ~\gev$) and a lighter one is the other doublet.
The decay width \cite{Lee:2013fda} is
\beq
\Gamma (H^\pm \to W h) \simeq \frac{\sin^2\beta}{16 \pi v^2} \frac{1}{m_{H^\pm}^3} \lambda^{3/2} (m_{H^\pm}^2, m_W^2, m_h^2)
\eeq
with $\lambda (x,y,z) \equiv x^2 + y^2 + z^2 - 2xy - 2yz -2zx$.
In this limit, we also get $\br (h \to Z' Z') \simeq 1$ as $h$ does not couple to the SM fermions.

In both scenarios (i) and (ii), for most of the parameter space, a dominant decay of the charged Higgs is into $Z'$s.
The dominance of the $H^\pm$ decay into the $Z'$ particles in both scenarios originates from the enhancement of couplings for boosted gauge bosons, formally known as Goldstone boson equivalence theorem.
Detailed discussions and illustrations are given in the aforementioned references \cite{Ramos:2013wea,Davoudiasl:2014mqa,Lee:2013fda}.
We will use a parameter for the charged Higgs to $Z'$s decay branching ratio
\beq
Y \equiv \br(H^\pm \to W + Z'\text{s}) ,
\eeq
which corresponds to $Y = \br(H^\pm \to W Z')$ with one $Z'$ [for scenario (i)], and $Y = \br(H^\pm \to W h) \, \br(h \to Z'Z')$ with two $Z'$s [for scenario (ii)].
In a large area of the parameter space, $Y \simeq 1$ is obtained.
In our analysis, instead of using an exact value which depends on various model parameters, we will take a range of $Y = 0.5 - 1$ (the branching ratio of $50 \%$ or greater).

\section{\boldmath Production and Decays of Top quark at the LHC}
\subsection{\boldmath Top pair production}
Top quarks can be produced in pair by dominant QCD process ($q \bar q'$, $g g$) or singly by electroweak process with a $W$ boson at the Tevatron and the LHC.
At the LHC, the $t \bar t$ can be produced abundantly via dominant gluon-gluon fusion.
The gluon fusion makes up $90 \%$ of the total $t \bar t$ for the center-of-mass energy $\sqrt{s} = 14 ~\tev$ (about $80 \%$ for $\sqrt{s} = 7 ~\tev$).

The total $t \bar t$ pair production cross section at the LHC (at next-to-next-to-leading order QCD corrections) is predicted to be $\sigma_{t \bar t} \simeq 167 ~\pb$ (for $7 ~\tev$), $239 ~\pb$ (for $8 ~\tev$), and $933 ~\pb$ (for $14 ~\tev$) \cite{Czakon:2013goa}.
Currently, at the LHC, integrated luminosity ($L$) is about $5 ~\fb^{-1}$ for $7 ~\tev$, and about $20 ~\fb^{-1}$ for $8 ~\tev$ \cite{ATLASluminosity}.
The $14 ~\tev$ LHC is scheduled to operate soon, and is expected to reach the luminosity of several hundreds of $\fb^{-1}$.

\begin{figure}[t]
\begin{center}
\includegraphics[width=0.45\textwidth,clip]{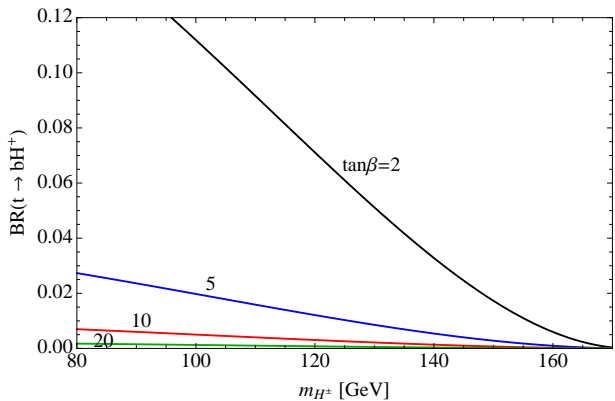}
\end{center}
\caption{$\br(t \to b H^+)$ for $\tan\beta = 2$ (black), $5$ (blue), $10$ (red), $20$ (green).
The $t$ decay into $H^+$ is larger for smaller $m_{H^\pm}$ and smaller $\tan\beta$.
}
\label{fig:tTobCH}
\end{figure}

\subsection{\boldmath Top decay into dark gauge bosons through $H^\pm$}
A top quark can decay into a charged Higgs through $t \to b H^+$ for a relatively light $H^\pm$.
While there are searches for light charged Higgs in $t \to b H^+ \to b \nu \tau^+$ and $t \to b H^+ \to b c \bar s$ based on typical 2HDMs, the results are negative \cite{Branco:2011iw}.
These searches, however, would have missed the light charged Higgs of $t \to b H^+ \to b W + Z'$s that would dominate in our scenario.
As described earlier, the $H^\pm$ can decay into the $Z'$s dominantly in the presence of the dark force.

Neglecting $m_b / m_t$ and higher order corrections, the relevant top decay widths are
\beq
\Gamma_{t \to b W} = \frac{\sqrt{2} G_F |V_{tb}|^2}{16 \pi} m_t^3 \left( 1 - \frac{m_W^2}{m_t^2} \right)^2 \left( 1 + \frac{2 m_W^2}{m_t^2} \right) ,
\label{eq:tbW}
\eeq
\beq
\Gamma_{t \to b H^+} = \frac{\sqrt{2} G_F |V_{tb}|^2}{16 \pi} m_t^3 \left( 1 - \frac{m_{H^\pm}^2}{m_t^2} \right)^2 \frac{1}{\tan^2\beta}
\label{eq:tbH}
\eeq
with $\tan\beta \gsim 1$ in the dark force model we consider.
(The $t \to b W$ decay itself shows the enhancement from the Goldstone boson equivalence theorem with a boosted $W$ boson \cite{Peskin}.)
As both decays have the same dependence on the Cabibbo-Kobayashi-Maskawa (CKM) matrix element $|V_{tb}|^2$, even quite sizable $\Gamma(t \to b H^+)$ may not alter an effective value of $V_{tb}$ significantly when it is measured from the top quark decays. 

In our study, we still take $t \to b W$ as the dominant top decay and consider $t \to b H^+$ as the important subdominant one.
Then the branching ratio of the on-shell $t \to b H^+$ decay, which is dependent on unknown $m_{H^\pm}$ and $\tan\beta$, is
\bea
\br(t \to b H^+) &\simeq& \frac{\Gamma_{t \to b H^+}}{\Gamma_{t \to b W} + \Gamma_{t \to b H^+}} \\
&\approx& \left( \frac{m_t^2 - m_{H^\pm}^2}{m_t^2 - m_W^2} \right)^2 \frac{1 / \tan^2\beta}{1 + 2 m_W^2 / m_t^2} , ~~~~ \label{eq:tTobCH}
\eea
which is plotted in Fig.~\ref{fig:tTobCH}.
We can see that a few $\%$ level of the top decay can be easily accommodated for $m_{H^\pm} < m_t$.
For $m_{H^\pm} = 140 ~\gev$, for instance, we have $\br(t \to b H^+) \simeq 0.03 - 0.0003$, with the $\tan\beta = 2 - 20$ range (higher BR corresponds to lower $\tan\beta$).

\section{\boldmath Setup for numerical anlaysis}
We will consider the $t \bar t$ production followed by one of the top quark pair decaying into $Z'$s.
We will also consider only the $Z' \to \ell^+ \ell^-$ channel.
The cross section at the LHC to produce $Z'$ is given by
\beq
\sigma(p p \to b W \, \bar b W + Z'\text{s}) \simeq \sigma_{t \bar t} \, 2 X
\eeq
using $\br (t \to b W) \simeq 1$ and a new parameter $X$ for the top to $Z'$ decay branching ratio
\beq
X \equiv \br(t \to b W + Z'\text{s}) , \label{eq:X}
\eeq
whose detail depends on the specific scenario.

In scenarios (i) and (ii), using the on-shell $H^\pm$ decays, we have
\beq
X = \br(t \to b H^+) \, Y
\eeq
with $\br(t \to b H^+)$ given in Fig.~\ref{fig:tTobCH}.
In the numerical analysis of this paper, we consider only the scenario (i).
[The application for the scenario (ii) would be straightforward.]
This process is
\beq
\bullet \text{ top pair: } p p \to t \bar t \to b W \bar b H^\pm \to b W \bar b W + Z' ,
\label{eq:topPair}
\eeq
which is similar to the dominant $t \bar t$ process except that $Z'$ is accompanied in the decay products.
We consider only $m_W \lsim m_{H^\pm} \lsim m_t$ to avoid additional constraints from $W \to b H^\pm$ decays, etc.

\begin{figure}[tb]
\begin{center}
\subfigure[]{
\includegraphics[width=0.45\textwidth,clip]{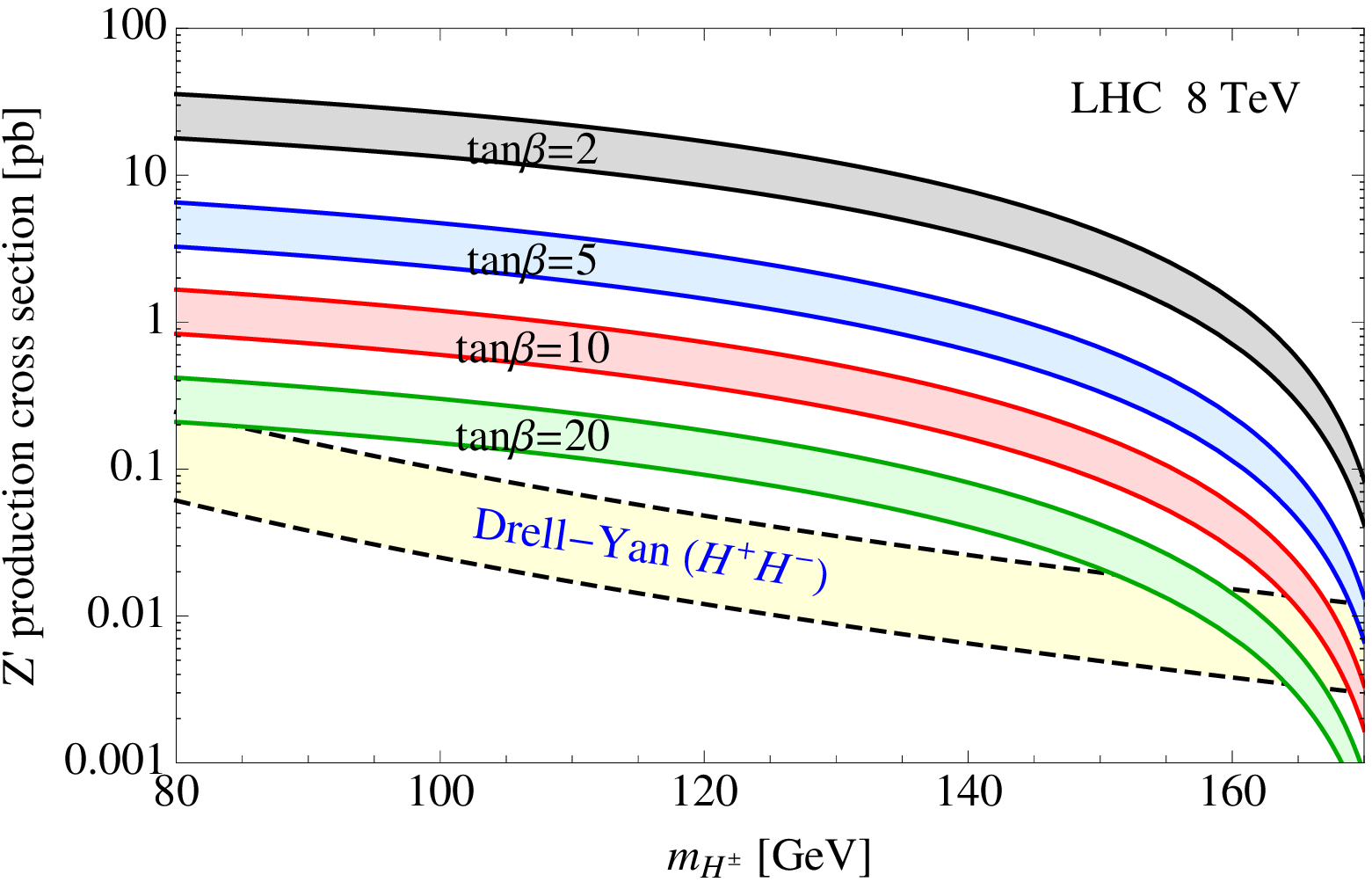}} \\
\subfigure[]{
\includegraphics[width=0.45\textwidth,clip]{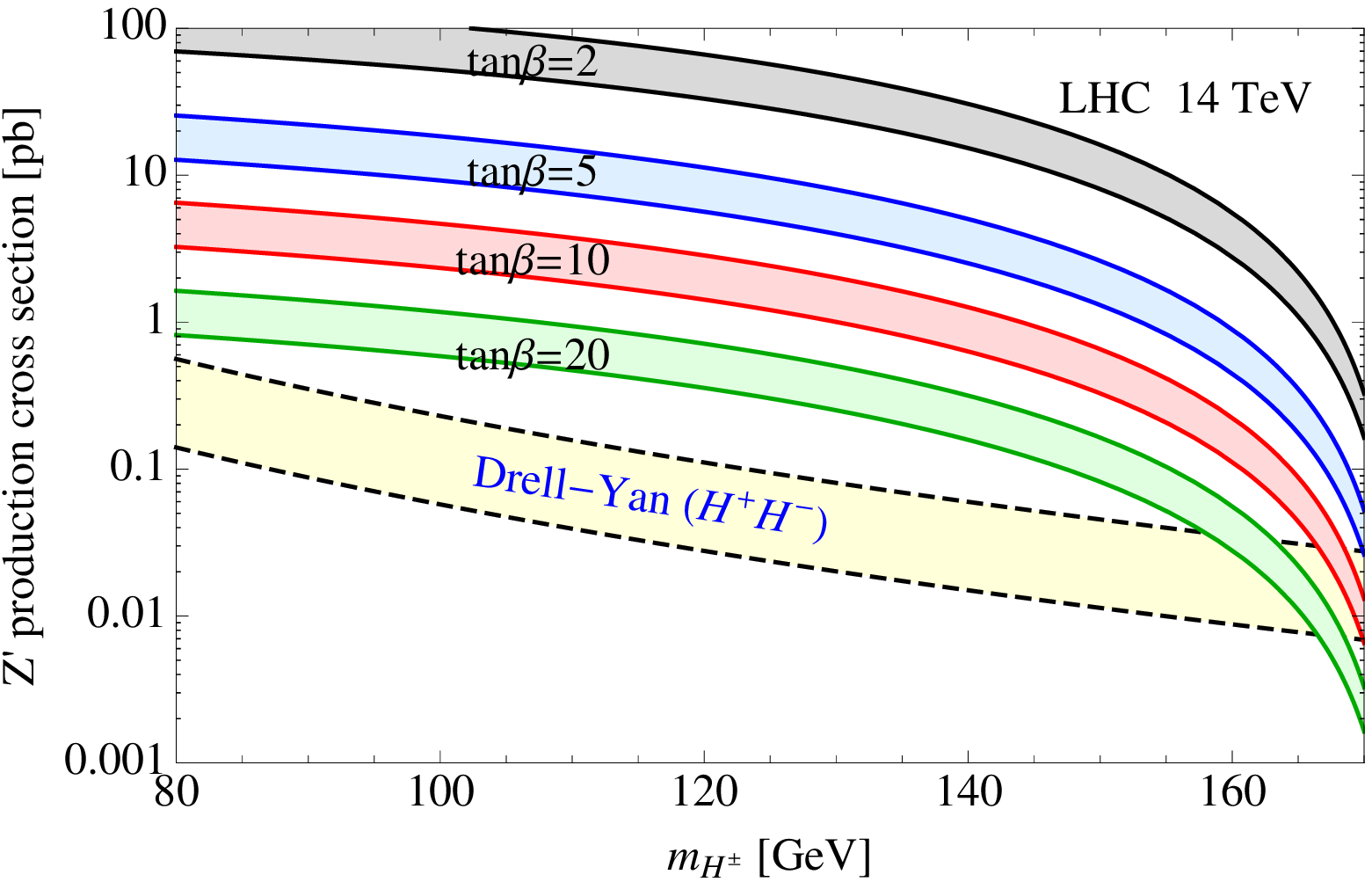}}
\end{center}
\caption{Production cross sections of the $Z'$ with $m_{H^\pm}$ in the $t \bar t$ channel ($p p \to t \bar t \to b W \bar b W + Z'$) at the (a) 8 TeV LHC and (b) 14 TeV LHC.
Cross section at the 14 TeV is about 4 times larger than that at the 8 TeV.
In the $m_W \lsim m_{H^\pm} \lsim m_t$ range, the results for $\tan\beta = 2$ (black), $5$ (blue), $10$ (red), $20$ (green) are shown.
Drell-Yan channel ($p p \to H^+ H^- \to W W + Z' Z'$) (Dashed) is also shown for comparison.
The band indicates $\br (H^\pm \to W Z') = 0.5 - 1$ range.
}
\label{fig:comparison}
\end{figure}

Figure~\ref{fig:comparison} shows the production cross sections of the $Z'$ with $m_{H^\pm}$ in the $t \bar t$ channel ($p p \to t \bar t \to b W \bar b W + Z'$) at the (a) 8 TeV LHC and (b) 14 TeV LHC, using the high order calculation of $\sigma(p p \to t \bar t)$.
The band indicates $Y = \br (H^\pm \to W Z') = 0.5 - 1$ range.

We will take a rather conservative value of $X = 0.001$ and assume $\br(Z' \to \ell^+ \ell^-) = 0.2$, for our illustration, meaning $0.02 \%$ of top quark decay is into the lepton pairs.
[Only the product $X \, \br(Z' \to \ell^+ \ell^-)$ has the real meaning for the final lepton pair production rate.]

The value $X = 0.001$ can be obtained, for example, for $m_{H^\pm} = 100 ~\gev$ (with $\tan\beta \simeq 20$, $Y \simeq 0.8$), $m_{H^\pm} = 140 ~\gev$ (with $\tan\beta \simeq 10$, $Y \simeq 0.9$), and $m_{H^\pm} = 160 ~\gev$ (with $\tan\beta \simeq 5$, $Y \simeq 1$).

We use a rather intuitive parameter $X$ for our presentation instead of model-specific parameters.
The exact value of $Y$, for example, can be more constrained for a specific $m_{H^\pm}$, $\tan\beta$, $m_{Z'}$.
There are many unknown parameters such as $\br(Z' \to \ell^+ \ell^-)$ and other dark sector related parameters anyway, and our treatment using $X$ (with a rather conservative values) allows us more controllable analysis.

\section{\boldmath Comparison to the charged Higgs pair production}
The charged Higgs boson can be typically produced in the Drell-Yan process.
Charged Higgs pair production through Drell-Yan ($\gamma^*$, $Z^*$) is model independent except for the mass.

The Drell-Yan production of $H^\pm$ pair, for scenario (i), is
\beq
\bullet \text{ Drell-Yan: } p p \to H^+ H^- \to W W + Z' Z' .
\eeq

In Fig.~\ref{fig:comparison}, we can see the difference of the production cross sections of the $Z'$ through the $t \bar t$ channel and Drell-Yan channel.
In both cases, the cross section decreases with $m_{H^\pm}$ because of the phase space.
We use the on-shell decays only using branching ratio of Fig.~\ref{fig:tTobCH}.

The Drell-Yan and the $t \bar t$ processes have a few differences.
First, the $t \bar t$ production cross section is much larger than the Drell-Yan case, except for the very large $\tan\beta$ ($\tan\beta \gsim 20$) or large $m_{H^\pm}$.
Second, $t \bar t$ produces only one charged Higgs while the Drell-Yan produces a pair of charged Higgs.
Third, tagging is different ($b W$ pair for $t \bar t$, $W$ pair for Drell-Yan).
With different tagging and a small production cross section, we neglect the Drell-Yan process in our study.

\section{\boldmath Discovery potential at the LHC}
\begin{figure}[tb]
\begin{center}
\subfigure[]{
\includegraphics[width=0.45\textwidth,clip]{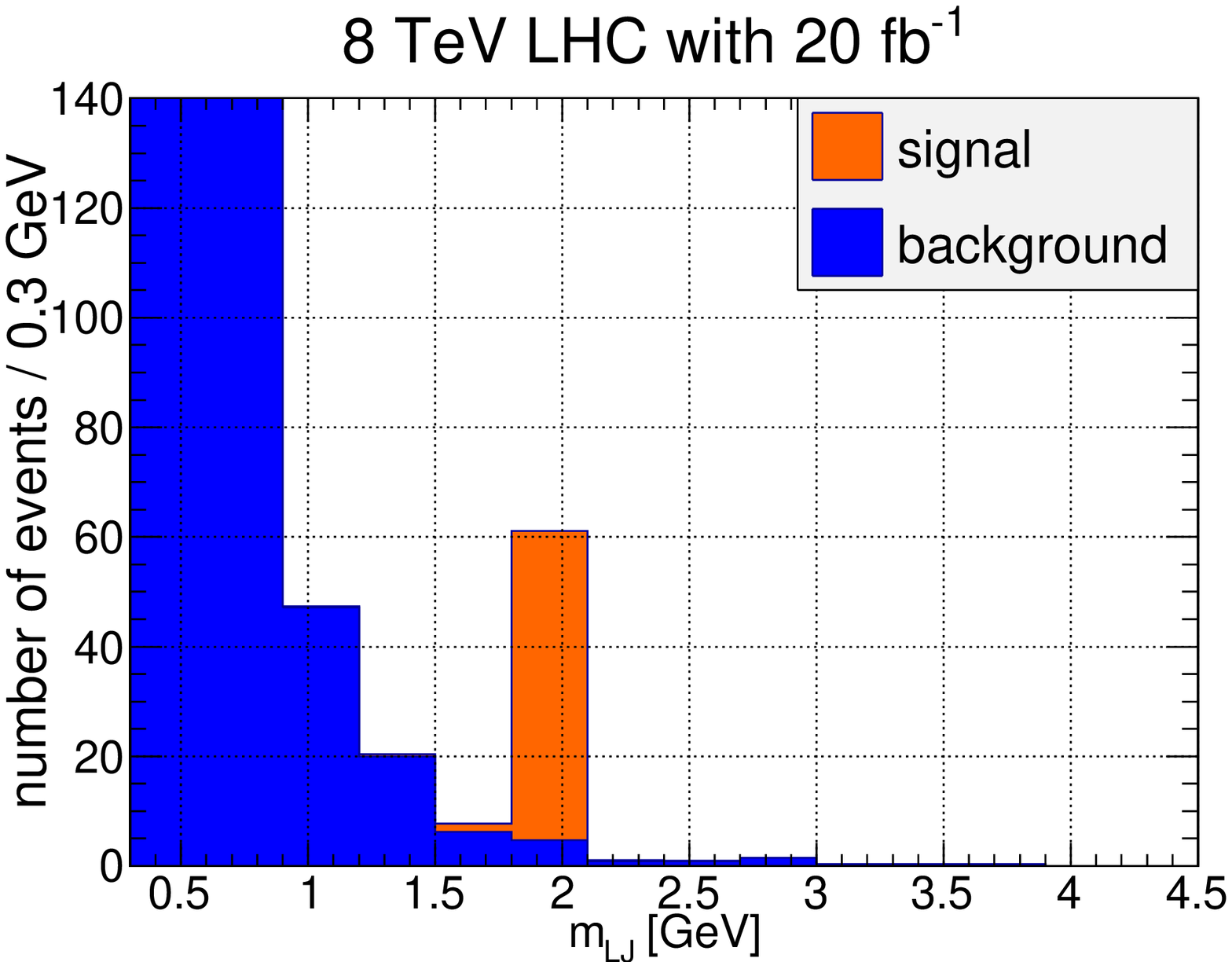}} \\
\subfigure[]{
\includegraphics[width=0.45\textwidth,clip]{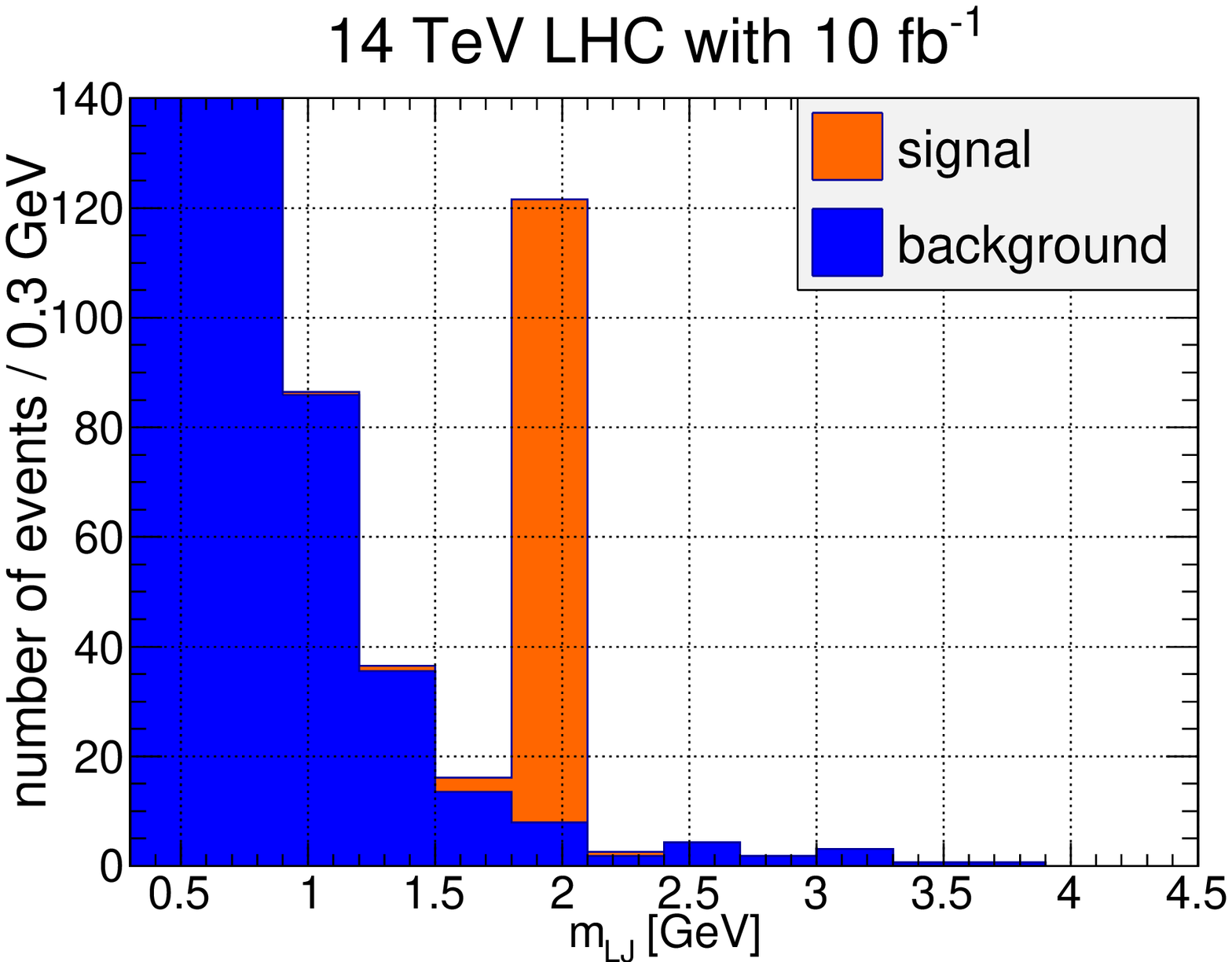}}
\end{center} 
\caption{Distributions of lepton-jet mass ($m_{\textrm{LJ}}$) in the $t \bar t + \text{LJ}$ mode at the LHC for (a) 8 TeV LHC with $L = 20 ~\fb^{-1}$ and (b) 14 TeV LHC with $L = 10 ~\fb^{-1}$. 
$m_{H^\pm} = 140 ~\gev$, $m_{Z'} = 2 ~\gev$ with $X = 0.001$ and $\br(Z' \to \ell^+ \ell^-) = 0.2 $ are used.
}
\label{fig:mljHIST}  
\end{figure}

We discuss the discovery potential of the $Z'$ from the $t \bar t$ process at the LHC.
For definiteness, we assume $\br(Z' \to \ell^+ \ell^-) = 0.2$.
For the background events, we take only the irreducible processes $t \bar t + \ell^+ \ell^-$ from the off-shell photon and $Z$ boson, although there may be more sources of backgrounds such as $t \bar t$ + jets with the jets faking leptons.
In this section, we require two $b$-tagged jets to limit backgrounds to $t\bar t$ events. 
The Appendix describes more details of our Monte Carlo study including the tagging efficiencies.
The efficiency depends on $m_{H^\pm}$ and $m_{Z'}$ as well as the LHC energy.
We assume a $K$-factor for the background events $K_{\textrm{bkg}}\simeq 2$ and apply it to our background events generated by tree-level Monte Carlo simulation.\footnote{The assumed $K_\text{bkg} \simeq 2$ can be compared to $K$ factors for signals $K_{\textrm{sig}} = 1.74$ $(1.84)$ for $8\,\tev\, (14\,\tev)$ LHC, obtained from the estimation with our leading order Monte Carlo simulation and Ref.~\cite{Czakon:2013goa}.}

First, we consider the $\sqrt{s} = 8\,\tev$ case ($\sigma_{t \bar t} \sim 239 ~\pb$), with integrated luminosity of about $L = 20 \,\fb^{-1}$.
Table~\ref{table:effS8TeV} shows the number of signal events and background events for various choices of $m_{H^\pm}$ and $m_{Z'}$, with $X = 0.001$ and $\br(Z' \to \ell^+ \ell^-) = 0.2$.
The numbers are for each lepton-jet bin ($20\%$ window of a given $m_{Z'}$) with two $b$-tagging. (See the Appendix for details including the tagging efficiencies for signal and background events.)

\begin{table}[t]
\centering
\begin{tabular}{|c||c|c|c|c|} 
\hline
$m_{Z'}$ & \multicolumn{3}{c|}{$m_{H^\pm}$} & \\
\hhline{~---~}
[GeV]& $100 ~\gev$ & $140 ~\gev$ & $160 ~\gev$ & BKG\\
\hhline{-----}
$1$ & 40.0 & 86.2  & 58.1  &  69.6 \\
$2$ & 8.2 & 59.9  & 47.8  & 5.0  \\
$5$ & 0.1 & 5.0 & 9.1  & 0.3   \\ 
\hline
\end{tabular}
\caption{\label{tab:cross} Expected number of events in each lepton-jet bin ($20\%$ window of the $Z'$ mass) with two $b$-tagging in $8 ~\tev$ LHC $20 ~\fb^{-1}$. We set $X=0.001$ and $\textrm{BR}(Z'\to\ell^+\ell^-)=0.2$.
Signal events were obtained with high order $\sigma_{t \bar t}$ with the branching ratio, and
the background events were obtained with tree-level simulation with $K_\text{bkg} = 2$.
\label{table:effS8TeV}
 }
\end{table}

For a specific example, we consider $m_{H^\pm} = 140\,\gev$ and $m_{Z'} = 2\,\gev$. 
We expect the number of signal events (after the tagging efficiency), for $L = 20\,\fb^{-1}$,
\beq
N_\text{sig} = \sigma_{t \bar t} \, 2 X \, \br(Z' \to \ell^+ \ell^-) \, \epsilon_\text{sig} \, L \\
\approx 60 ,
\eeq
and the SM background events
\beq
N_\text{bkg} = \sigma_\text{bkg} \, \epsilon_\text{bkg} \, L \\
\approx 5 ,
\eeq
which would give a very large number of signal events compared to the backgrounds (resulting the likelihood ratio $S_{\textrm{cL}} \simeq 14.6$).
The likelihood ratio, which is defined as
\beq
S_{\textrm{cL}} = \sqrt{2 N_\textrm{obs} \log\left(1+N_\textrm{sig}/N_{\textrm{bkg}}\right)-2 N_{\textrm{sig}}}
\eeq
with $N_\textrm{obs} = N_\textrm{sig} + N_\textrm{bkg}$, is a good method even when there are relatively small background events.

Even this kind of large signal can be still missed in conventional analysis.
For example, in the CMS $t \bar t$ dilepton analysis (see the Appendix) with $5.3\,\fb^{-1}$ luminosity at $8\,\tev$ LHC \cite{Chatrchyan:2013faa}, signals can be lost by invariant mass requirements for the lepton pair ($m_{\ell\ell} > 20 ~\gev$).   
The contribution of signals to dileptons in the $t\bar t$ channels are very small with CMS analysis cuts (one $b$-tagged jet). For the above sample point, we estimate the expected number of signal, $N_{\textrm{sig}} \simeq 4$ ($\epsilon_{\textrm{sig}} \simeq 0.71\%$) can be buried under the uncertainties of $t\bar t$ dilepton events (Expected uncertainty in $t\bar t$ dilepton samples, $\Delta N_{\textrm{bkg}} \simeq 591$ with observed data $N_{\textrm{bkg}} \simeq 1.7\times 10^4$ ), resulting in only $S_{\textrm{cL}} \simeq 0.03$.

Thus, re-analysis of existing 8 TeV data with $t \bar t$ + lepton-jet can potentially bring a discovery of the light $Z'$.

For the $\sqrt{s} = 14 ~\tev$ case ($\sigma_{t \bar t} \sim 933 ~\pb$), we show the required luminosity for $S_{\textrm{cL}} =5$ (corresponding to $5 \sigma$ discovery) in Table\,\ref{table:effS14TeV}.
Basically the same method as the $8 ~\tev$ case (Table~\ref{table:effS8TeV}) is used.
It shows the dark force search at the very early stage of the 14 TeV LHC will be a very interesting program.
Both Table~\ref{table:effS8TeV} and Table~\ref{table:effS14TeV}, for given $X$ and $\br(Z' \to \ell^+ \ell^-)$, can be obtained from Table~\ref{table:effS} in the Appendix up to the precision.

Figure~\ref{fig:mljHIST} shows the signals and backgrounds for the above sample point ($m_{H^\pm} = 140 ~\gev$, $m_{Z'} = 2 ~\gev$, $X = 0.001$) with $\br(Z' \to \ell^+ \ell^-) = 0.2$.
For cuts, we require the CMS-like analysis cuts\,[Sec.~\ref{sec:cut}] with two $b$-tagged jets and construct lepton-jets with $m_{\textrm{LJ}}>0.2 ~\gev$. For backgrounds, the leading order cross section is $3.73\,\textrm{pb}$ at $8\, \tev$ and $15.33\,\textrm{pb}$ at $14 ~\tev$ LHC with parton level cuts for the rapidity of quark and leptons as $|\eta_q|<5$, $|\eta_{\ell}|<3$ and a dilepton invariant mass cut for the same flavor, an opposite-sign lepton pair greater than $0.2\,\gev$. Corresponding efficiencies are about $1.4\%$ for both collision energies. We take $K_\text{bkg} = 2$ for the $K$-factor of the background. 
We can see that the signal shows up as a clear spike over the SM background.

The $t \to b W + Z'$ decay (via $H^\pm$) can be compared to the dominant top decay mode $t \to b W$.
The top decay into the charged Higgs might look similar to the dominant top decay accompanied by a $Z'$ that can decay into a pair of collimated leptons or others that are hard to identify.
It has important implication as the current experimental measurements might have counted the $t \to b W + Z'$ as $t \to b W$ depending on the analysis methods.
$Z'$ may also have sizable decay branching ratio into neutrinos or invisible particles.
As mentioned earlier, since both decays have the same dependence on the CKM matrix element $|V_{tb}|^2$, [Eqs.~\eqref{eq:tbW} and \eqref{eq:tbH}], even quite sizable $\Gamma(t \to b H^+)$ may not alter the effective value of $V_{tb}$ significantly when it is measured from the top quark decays.

\begin{table}[t]
\centering
\begin{tabular}{|c||c|c|c|} 
\hline
 $m_{Z'}$ & \multicolumn{3}{c|}{$m_{H^\pm}$}  \\
\hhline{~---}
[GeV]& $100 ~\gev$ & $140 ~\gev$ & $160 ~\gev$\\
\hhline{----}
$1$ & $7.8\,\fb^{-1}$ & $1.9\,\fb^{-1}$  & $3.4\,\fb^{-1}$ \\
$2$ & $14.5\,\fb^{-1}$ & $0.7\,\fb^{-1}$ & $1.0\,\fb^{-1}$  \\
$5$ & - & $7.3\,\fb^{-1}$  & $3.5\,\fb^{-1}$   \\ 
\hline
\end{tabular}
\caption{\label{tab:cross} Required luminosity for $14 ~\tev$ LHC to see the likelihood ratio $S_\textrm{cL} = 5$ (corresponding to $5 \sigma$ discovery).
Basically the same method as Table~\ref{table:effS8TeV} is used.
\label{table:effS14TeV}
 }
\end{table}

\section{Summary}
We discussed the production of light $Z'$ gauge boson through a top quark at the LHC.
A light $Z'$ of roughly ${\cal O} (1) ~\gev$ with small coupling is a very well motivated new physics candidate as it can address some astrophysical anomalies as well as the muon $g-2$ anomaly.
While its search is very active at the low energy experimental facilities, its search at the LHC is relatively limited so far.

The LHC can produce the top quark pair abundantly through the gluon fusion.
We considered the scenario the top quark decays through a charged Higgs $t \to b H^+$, where the charged Higgs can decay into one $Z'$ or multiple $Z'$s dominantly.
The top decay into the dark gauge boson is very close to its dominant decay mode accompanied by one or two illusive $Z'$s ($t \to b W + Z'$s).
Even a small $\br (t \to b H^+)$ can be enough to produce the $Z'$ at the observable level at the LHC experiments.
The $Z'$ production through a top can be larger than the typical Drell-Yan mechanism to produce a pair of charged Higgs and produce one $Z'$ when the Drell-Yan produces a pair of $Z'$.

We considered the process $p p \to t \bar t \to b W \bar b W + Z'$ with $Z'$ decaying into a lepton pair.
Because of lightness of the $Z'$, the lepton pairs are highly collimated forming lepton-jets.
For some parameter space, even the existing $8 ~\tev$ LHC data may give enough signals for a discovery.
It also guarantees a huge discovery potential even at the very early stage of the $14 ~\tev$ LHC experiments.

As the top quark decaying into the dark sector mode can be easily mistaken with its dominant $b W$ mode, a reanalysis of the top data at the hadron collider can possibly reveal an interesting hint of the $Z'$ even if the $Z'$ is very elusive or decays invisibly.
It calls for attention in both the experimental and theoretical sides of the top quark study.

Note: Around the time this work was submitted, Ref.~\cite{Davoudiasl:2014mqa} preprint (version 2), which overlaps with some aspect of this paper, appeared.

\acknowledgments
K.K. is supported by the U.S. DOE under Grant No.~DE-FG02-12ER41809, the University of Kansas General Research Fund allocation 2301566.
H.L. is supported by U.S. DOE under Grant No.~DE-AC05-06OR23177, the NSF under Grant No.~PHY-1068008.
M.P. is supported by the World Premier International Research Center Initiative (WPI Initiative), MEXT, Japan.

\begin{figure*}[t] 
\begin{center}
\subfigure[]{
\includegraphics[width=0.35\textwidth,clip]{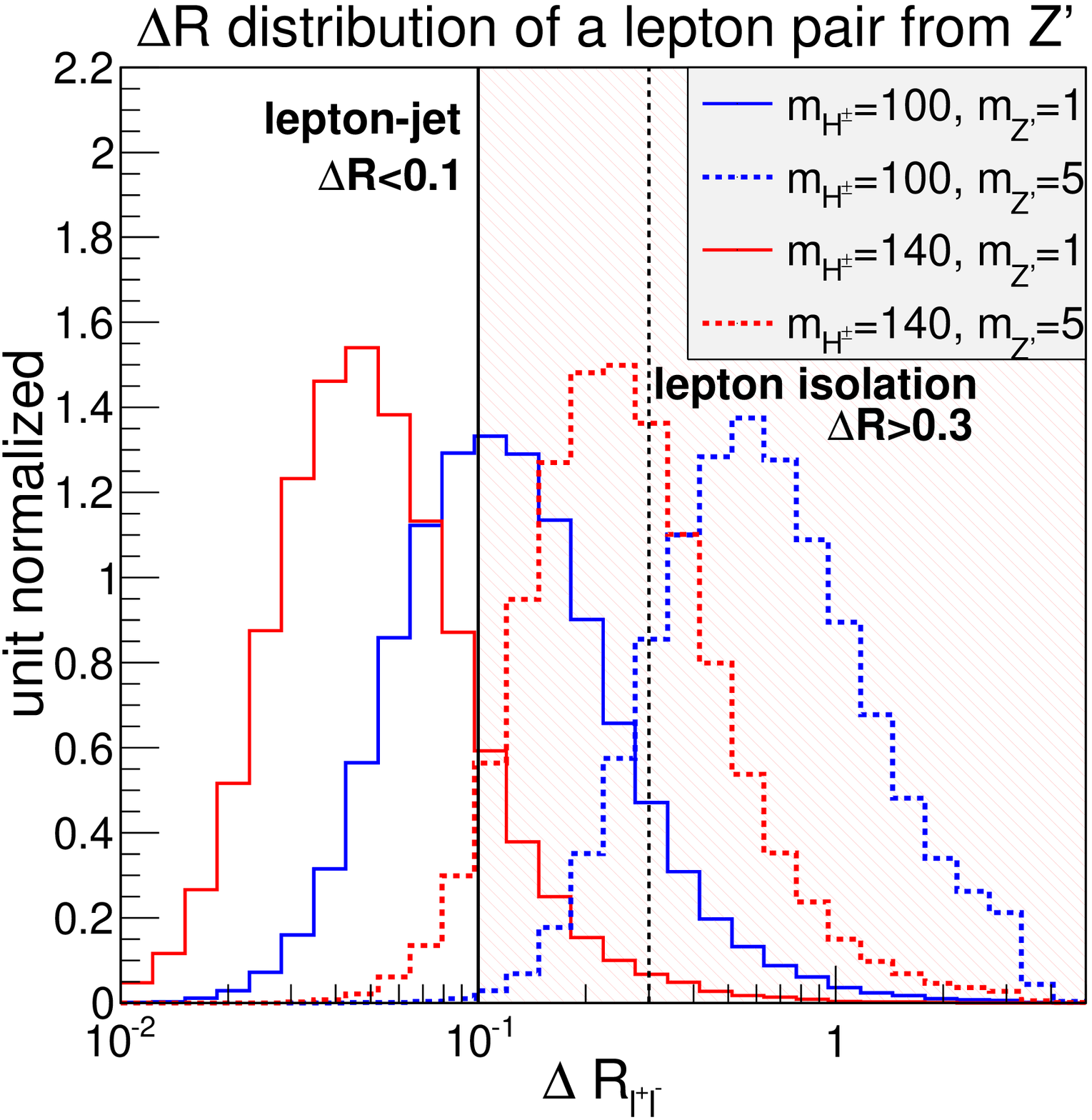}} ~~~
\subfigure[]{
\includegraphics[width=0.35\textwidth,clip]{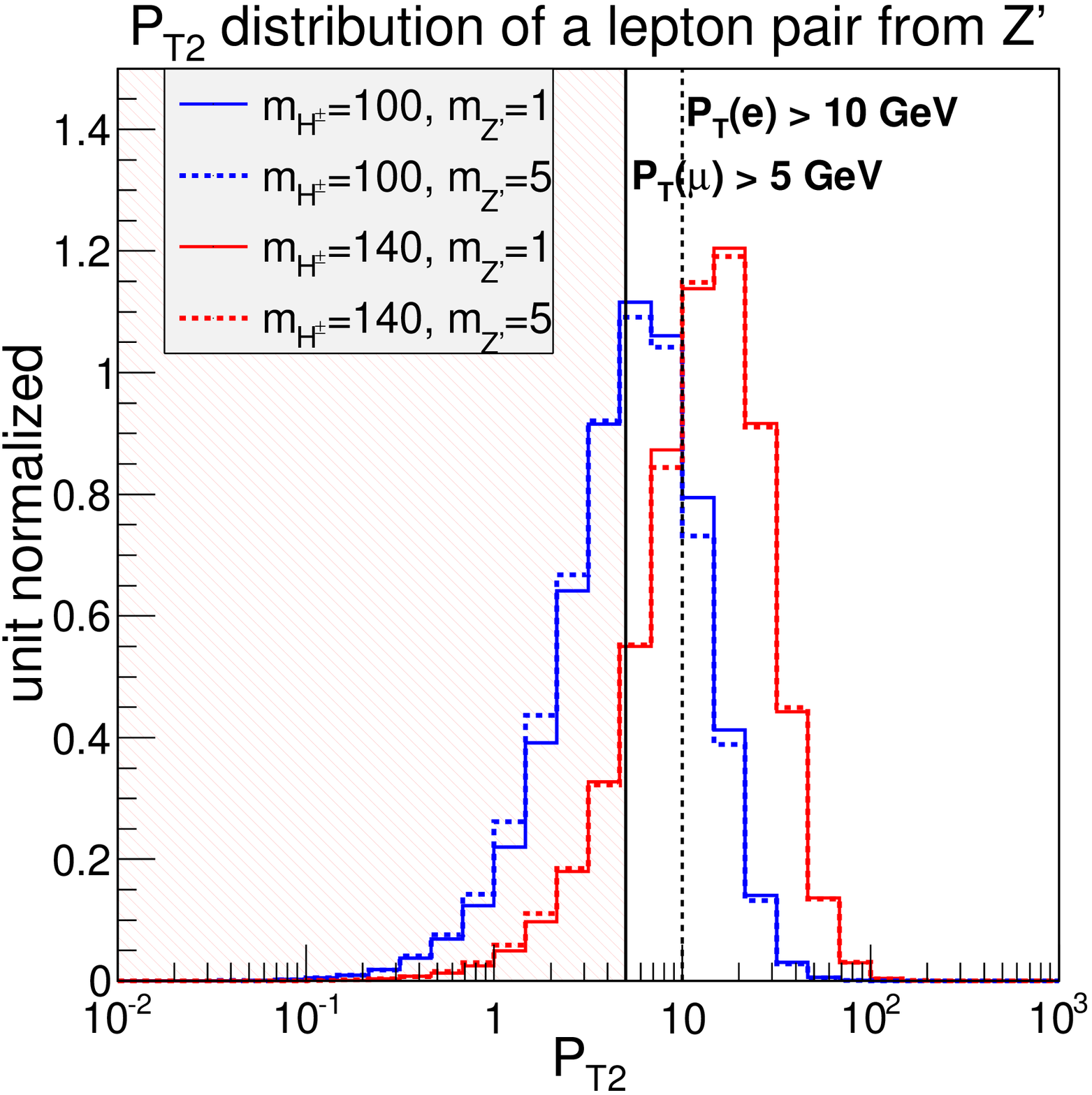}}
\caption{We show (a) $\Delta R$ and (b) $P_{T2}=\min[{P_{T(\ell)}, P_{T(\bar\ell)}]}$ of a lepton pair from the $Z'$ decay at the parton level for the $14 ~\tev$ LHC, which are useful to understand lepton-jet tagging efficiency.} 
\label{fig:why}
\end{center}
\end{figure*}

\appendix
\section{Some details on numerical analysis}
\subsection{\boldmath Lepton pair from $Z'$ decay}
Light $Z'$ cannot be reconstructed with the usual lepton tagging because of the following simple kinematical reason.
The invariant mass of a lepton pair can be expressed as
\bea
m_{\ell^+\ell^-}^2 &=& 2 P_{T_1} P_{T_2} \left(\cosh{\Delta \eta}-1 \right) \\
&\simeq& 2 P_{T_1} P_{T_2} \left(\cosh{\Delta R}-1 \right) ,
\label{eq:mll}
\eea
with observation $\Delta R \simeq \Delta \eta$ since $\Delta \phi$ distribution is peaked at 0.
For a moderate lepton tagging efficiency, most analyses require $P_{T(\ell)}^{\textrm{min}}$ as a $P_T$ cut of leptons
\beq
P_{T(e)}^{\textrm{min}} = 10 ~\gev , \quad P_{T(\mu)}^{\textrm{min}} = 5 ~\gev .
\eeq
Now with an isolation requirement of $\Delta R > 0.3$, the corresponding minimum invariant mass of an electron pair and muon pair from $Z'$ would be
\bea
m_{e e} > \sqrt{2P_{T(e)}^{\textrm{min}} P_{T(e)}^{\textrm{min}} (\cosh(0.3)-1)} \simeq 3 ~\gev, \\
m_{\mu \mu} > \sqrt{2P_{T(\mu)}^{\textrm{min}} P_{T(\mu)}^{\textrm{min}} (\cosh(0.3)-1)} \simeq 1.5 ~\gev. 
\eea
Therefore conventional analyses would miss $Z'$ lighter than 3 (1.5) GeV in the dielectron (dimuon) channel.
We adopt an analysis method with lepton-jet proposed in Ref.~\cite{ArkaniHamed:2008qp}. 
A variation of the LJ definition can be found in Ref.~\cite{Aad:2013yqp}.
Since $t \bar t + \ell^+ \ell^-$ is a major background in our study, we follow the LJ definition in Ref.~\cite{Cheung:2009su} with a modification to the muon $P_T$ requirement.
\begin{itemize}
\item[1.] At least two same flavor leptons with $P_T > 10 ~\gev$ (electron), $5 ~\gev$ (muon) and in a cone of $\Delta R<0.1$.
\item[2.] Isolation: Hadronic and leptonic isolation of $\sum P_T < 3 ~\gev$ in $0.1<\Delta R<0.4$. 
\end{itemize}
In addition, to reduce the background events, we require $20\%$ window of the expected $Z'$ mass for an invariant mass of lepton-jet.
\begin{itemize}
\item[3.] Invariant mass cut on the lepton-jet: $|m_{\textrm{LJ}} -m_{Z'}| < 0.2\times m_{Z'}$.  
\end{itemize}

For a Monte Carlo simulation, we add $Z'$ and $H^\pm$ to the SM using FeynRules v2\,\cite{Alloul:2013bka} and simulate events with Madgraph v5.14\,\cite{Alwall:2011uj}, Pythia 6\,\cite{Sjostrand:2006za} and Delphes 3.0.11\,\cite{deFavereau:2013fsa}.
We modify $b$-jet tagging/mistagging efficiency (tagging efficiency is around $60\% - 75\%$ depending on $P_T$ and $\eta$) according to CMS CSVM tagging \cite{btagging,Chatrchyan:2012jua}.
We make a change in the Delphes module for a smearing muon $P_T$ to reflect the nonzero muon mass in the muon's four vector.
For the lepton-jet analysis, we add the lepton-jet class to the Delphes detector simulation.

\subsection{Signal boxes}
\label{sec:cut}
To consider all $t\bar t$ decay modes, we consider three signal boxes with a slight modification of CMS analyses according to the number of triggered leptons. Jets are reconstructed with an anti-$k_T$ algorithm with $\Delta R < 0.5$. 
We require one $b$-tagged jet (or two $b$-tagged jets) and the aforementioned lepton-jet mass window.
Other event selection criteria will depend on signal boxes as following.

\subsubsection{Dilepton box}
For event selections, we require the following criteria \cite{Chatrchyan:2013faa}. 
\begin{itemize}
\item[1.] Electrons with $P_T > 20 ~\gev,  |\eta|<2.5$, muons with $P_T > 20 ~\gev, |\eta|<2.1$ are taken.
\item[2.] For the same flavor opposite-sign leptons, we veto events with the following invariant mass window: $m_{\ell^+\ell^-} < 20 ~\gev$ and $|m_Z - m_{\ell^+\ell^-}| < 15 ~\gev$. 
\item[3.] For the same flavor lepton pairing, we require $\met > 40 ~\gev$.
\item[4.] We require at least two jets with $P_T > 30 ~\gev$, $|\eta|<2.5$.
\end{itemize}

\subsubsection{Semilepton box}
If events can not be put in the dilepton box, we pass them to the semilepton box with the following triggers.
We collect leptons of $P_T > 17 ~\gev$ if the lepton is a muon and of $P_T > 25 ~\gev$ for electrons as a lepton pool, then we require events to have exactly one lepton with the following selection criteria \cite{CMS:semilep}.
\bea
\textrm{muon} :&& P_T > 26 ~\gev,~ |\eta| < 2.1 , \nonumber \\
\textrm{electron} :&& P_T > 30 ~\gev,~ |\eta| < 2.5 .
\eea
For jets, we require at least four jets with $P_{T1}$, $P_{T2} > 45 ~\gev$ and $P_{T3}$, $P_{T4} > 35 ~\gev$.

\subsubsection{Hadronic top box}
If events are not included in the dilepton or semilepton boxes, we pass events to a hadronic top box where we give conditions only to jets.
Since we tag a lepton-jet, we relax the $P_T$ conditions of jets compared to the CMS analysis \cite{Chatrchyan:2013xza}. We require at least six jets with $P_{Tj} > 30 ~\gev$ with $|\eta|<2.4$.
The CMS requires $P_{T1},\,P_{T2},\,P_{T3},\,P_{T4} > 60 ~\gev$, $P_{T5} > 50 ~\gev$, $P_{T4} > 30\,\gev$, and additional constraints for the distance between two $b$-tagged jets and a kinematic fit for the mass reconstruction of $t$, $\bar t$, and $W$.

\begin{table*}[tb]
\centering
\begin{tabular}{|c|c||c|c|c||c|c|c|} 
\hline
LHC &$m_{Z'}$& \multicolumn{3}{c||}{$\epsilon_\textrm{LJ} (\epsilon_\textrm{(LJ+CMS)})$ [\%] for signal} & Mass range of & $\sigma_\text{bkg}^\text{LO}$  & $\epsilon_\textrm{LJ} (\epsilon_\textrm{(LJ+CMS)})$ [\%] \\
\hhline{~~---}
[$\tev$] & [$\gev$]& $m_{H^\pm} = 100 ~\gev$ & $m_{H^\pm} = 140 ~\gev$ & $m_{H^\pm} = 160 ~\gev$ & $ m_{\ell^+ \ell^-}\, [\gev]$ & [pb] & for background  \\
\hhline{--------}
\multirow{ 3}{*}{$8$ } & $1$ & 16.37 (4.18/2.07)  &  46.77 (10.96/4.51) & 52.04 (9.40/3.04)&	$0.5 - 1.5$ & $0.617$ & 2.05 (0.61/0.28)    \\
& $2$ &  3.07 (0.92/0.43) &  31.01 (7.64/3.13)  &   40.74 (7.57/2.50) &	$1.0 - 3.0$  & $0.157$  & 0.53 (0.19/0.08)  \\
& $5$  & 0.02 (0.00/0.00)  & 2.24 (0.64/0.26)  &   5.55 (1.25/0.48) &	$3.0 - 5.0$  & $0.0175$    & 0.32 (0.10/0.04) \\
\hhline{--------}\hhline{--------}
\multirow{ 3}{*}{$14$ } &$1$ & 16.38 (4.28/2.02)  &  44.28 (10.73/4.37) & 50.54 (9.44/3.13)  &	$0.5 - 1.5$ & $2.536$ & 2.18 (0.60/0.30)    \\
& $2$ &  3.33 (1.11/0.49) &  29.73 (7.52/3.13)  &   39.31 (7.64/2.51)  &	$1.0 - 3.0$ & $0.640$  & 0.57 (0.23/0.11)  \\
& $5$  & 0.03 (0.01/0.00)  & 2.57 (0.76/0.28)  &   5.90 (1.40/0.47)  &	$3.0 - 5.0$ & $0.0706$    & 0.34 (0.15/0.08) \\
\hline
\end{tabular}
\caption{\label{tab:cross} Lepton-jet tagging efficiency $\epsilon_\textrm{LJ}$ ($\%$) in $p p \to b W \bar b W + \ell^+ \ell^-$ for signal (for given $m_{H^\pm}$ and $m_{Z'}$) and background (from virtual photon and virtual $Z$ boson) at the 8 and 14 TeV LHC.
The numbers in parentheses ($\epsilon_\textrm{(LJ+CMS[1b])} / \epsilon_\textrm{(LJ+CMS[2b])}$) are the efficiencies when we require additional selection cuts, 
 requiring one $b$-tagged or two $b$-tagged jets as described in Appendix~\ref{sec:cut}.
Coupling structure of $Z'$ to the lepton does not give a significant effect on the tagging efficiency. In the above table, we take axial coupling as an example. 
For backgrounds, we set the trigger of a $m_{\ell^+ \ell^-}$ mass window as in the table to enlarge statistics.
\label{table:effS}
 }
\end{table*}

\subsubsection{Backgrounds}
We consider only the irreducible background of $(t \bar t + \ell^+ \ell^-)$ from the virtual photon and virtual $Z$ boson radiations and we do not consider possible reducible backgrounds from the mistagged lepton-jet.
For example, jets can be misidentified as electrons and overlapped leptons as discussed in Ref.~\cite{Aad:2012qua}.

\subsection{Simple explanation of a lepton-jet tagging efficiency for signal events}
With a good approximation, most $t\bar t$ pairs will be produced near the energy threshold. 
At the $t(\bar t)$ rest frame, 
the energy distribution of a lepton from $Z'$ will drop logarithmically (from flat distribution) \cite{Agashe:2012fs} after 
\beq
E_\ell^{(\textrm{cusp})} \equiv \frac{m_{Z'}}{2} e^{|\eta_{Z'}-\eta_{H^\pm}|},
\label{eq:peak}
\eeq
until
\beq
E_\ell^{(\max)} = \frac{m_{Z'}}{2} e^{(\eta_{Z'}+\eta_{H^\pm})},
\eeq
with boost factors (rapidity) of $H^\pm$ and $Z'$ as
\bea
\eta_{H^\pm} &=& \cosh^{-1}\left({\frac{m_t^2 + m_{H^\pm}^2-m_b^2}{2 m_t m_{H^\pm}}}\right) , \\
\eta_{Z'} &=& \cosh^{-1}\left({\frac{m_{H^\pm}^2+m_{Z'}^2-m_W^2}{2  m_{Z'} m_{H^\pm}}}\right). 
\eea

Thus if we assume that a distribution of the geometric average of leptons' $P_T$ is localized around $P_T^{\textrm{peak}}\equiv \frac{1}{2} E_\ell^{(\textrm{cusp})}$, 
we can estimate the peak of $\Delta R\,(=\Delta R^{\textrm{(peak)}})$ between two leptons from $Z'$ decays with Eq.~\eqref{eq:mll} by
\beq
\Delta R^{\textrm{(peak)}} \sim \cosh^{-1}\left(\frac{2m_{Z'}^2}{(E_\ell^{(\textrm{cusp})})^2}+1\right).
\label{eq:deltaR}
\eeq

With this simple kinematical study, we can explain why $\Delta R$ will be the major criteria for a lepton-jet tagging. 
For example, with $m_{H^\pm}=140\,\gev$, $P_{T}$ of the second hardest lepton ($P_{T2}$) does not change much from $m_{Z'}= 1$ to $5 ~\gev$ (Fig.~\ref{fig:why}).
This can be understood since $P_T^{\textrm{peak}}$ does not change much from $18.96$ to $18.95\,\gev$ with Eq.\,\eqref{eq:peak}. 
But the corresponding $\Delta R^{\textrm{(peak)}}$ will change significantly to the point to change lepton-jet tagging efficiency from $\Delta R^{\textrm{(peak)}}\sim 0.05$ to $0.26$ estimated by Eq.~\eqref{eq:deltaR}.
The observed peak is similar to this estimation as in Fig.~\ref{fig:why}.
(For $m_{H^\pm} = 100\,\gev$, estimated $\Delta R^{\textrm{(peak)}}$ is over estimated by a factor of 2 since the actual $P_T$ of leptons is not very well localized.)
Thus, for a large $m_{Z'}$, lepton-jet tagging efficiency is low due to large $\Delta R$ between leptons from $Z'$. 

Now we consider the effect from the mass of charged Higgs.
The $P_T$ of leptons increases with $m_{H^\pm}$. At the same time, due to the dilepton mass relation, Eq.\,(\ref{eq:mll}), $\Delta R$ decreases with $m_{H^\pm}$. 
This effect from $\Delta R$ is greater than the effect from $P_T$, and the lepton-jet tagging efficiency increases with $m_{H^\pm}$.




\begin{thebibliography}{99}
\bibitem{Aad:2012tfa} 
  G.~Aad {\it et al.}  [ATLAS Collaboration],
  Phys.\ Lett.\ B {\bf 716}, 1 (2012)
  [arXiv:1207.7214 [hep-ex]].

\bibitem{Chatrchyan:2012ufa} 
  S.~Chatrchyan {\it et al.}  [CMS Collaboration],
  Phys.\ Lett.\ B {\bf 716}, 30 (2012)
  [arXiv:1207.7235 [hep-ex]].
  
\bibitem{PDG} 
  J.~Beringer {\it et al.}  [Particle Data Group Collaboration],
  Phys.\ Rev.\ D {\bf 86}, 010001 (2012)
  and 2013 partial update for the 2014 edition \url{http://pdg.lbl.gov}.
  
\bibitem{ArkaniHamed:2008qn} 
  For example, see 
  N.~Arkani-Hamed, D.~P.~Finkbeiner, T.~R.~Slatyer and N.~Weiner,
  Phys.\ Rev.\ D {\bf 79}, 015014 (2009)
  [arXiv:0810.0713 [hep-ph]].

\bibitem{Adriani:2008zr} 
  O.~Adriani {\it et al.}  [PAMELA Collaboration],
  Nature {\bf 458}, 607 (2009)
  [arXiv:0810.4995 [astro-ph]].

\bibitem{Aguilar:2013qda} 
  M.~Aguilar {\it et al.}  [AMS Collaboration],
  Phys.\ Rev.\ Lett.\  {\bf 110}, 141102 (2013).

\bibitem{Fayet:2007ua} 
  P.~Fayet,
  Phys.\ Rev.\ D {\bf 75}, 115017 (2007)
  [hep-ph/0702176 [HEP-PH]].
  
\bibitem{Pospelov:2008zw} 
  M.~Pospelov,
  Phys.\ Rev.\ D {\bf 80}, 095002 (2009)
  [arXiv:0811.1030 [hep-ph]].
    
\bibitem{Essig:2013lka} 
  R.~Essig, J.~A.~Jaros, W.~Wester, P.~H.~Adrian, S.~Andreas, T.~Averett, O.~Baker and B.~Batell {\it et al.},
  arXiv:1311.0029 [hep-ph].

\bibitem{Bjorken:2009mm} 
  J.~D.~Bjorken, R.~Essig, P.~Schuster and N.~Toro,
  Phys.\ Rev.\ D {\bf 80}, 075018 (2009)
  [arXiv:0906.0580 [hep-ph]].

\bibitem{ArkaniHamed:2008qp} 
  N.~Arkani-Hamed and N.~Weiner,
  JHEP {\bf 0812}, 104 (2008)
  [arXiv:0810.0714 [hep-ph]].

\bibitem{Davoudiasl:2012ag} 
  H.~Davoudiasl, H.-S.~Lee and W.~J.~Marciano,
  Phys.\ Rev.\ D {\bf 85}, 115019 (2012)
  [arXiv:1203.2947 [hep-ph]].

\bibitem{Davoudiasl:2013aya} 
  H.~Davoudiasl, H.-S.~Lee, I.~Lewis and W.~J.~Marciano,
  Phys.\  Rev.\  D 88, {\bf 015022} (2013)
  [arXiv:1304.4935 [hep-ph]].

\bibitem{Lee:2013fda} 
  H.-S.~Lee and M.~Sher,
  Phys.\ Rev.\ D {\bf 87}, 115009 (2013)
  [arXiv:1303.6653 [hep-ph]].

\bibitem{Bigi:1986jk} 
  I.~I.~Y.~Bigi, Y.~L.~Dokshitzer, V.~A.~Khoze, J.~H.~Kuhn and P.~M.~Zerwas,
  Phys.\ Lett.\ B {\bf 181}, 157 (1986).
    
\bibitem{Abe:1995hr} 
  F.~Abe {\it et al.}  [CDF Collaboration],
  Phys.\ Rev.\ Lett.\  {\bf 74}, 2626 (1995)
  [hep-ex/9503002].
  
\bibitem{Abachi:1995iq} 
  S.~Abachi {\it et al.}  [D0 Collaboration],
  Phys.\ Rev.\ Lett.\  {\bf 74}, 2632 (1995)
  [hep-ex/9503003].

\bibitem{Ruegg:2003ps} 
  For a review, see H.~Ruegg and M.~Ruiz-Altaba,
  Int.\ J.\ Mod.\ Phys.\ A {\bf 19}, 3265 (2004)
  [hep-th/0304245].

\bibitem{Holdom:1985ag} 
  B.~Holdom,
  Phys.\ Lett.\ B {\bf 166}, 196 (1986).

\bibitem{Davoudiasl:2012qa} 
  H.~Davoudiasl, H.-S.~Lee and W.~J.~Marciano,
  Phys.\ Rev.\ Lett.\  {\bf 109}, 031802 (2012)
  [arXiv:1205.2709 [hep-ph]].

\bibitem{Langacker:2008yv} 
  P.~Langacker,
  Rev.\ Mod.\ Phys.\  {\bf 81}, 1199 (2009)
  [arXiv:0801.1345 [hep-ph]].

\bibitem{Alves:2013tqa} 
  A.~Alves, S.~Profumo and F.~S.~Queiroz,
ÊÊarXiv:1312.5281 [hep-ph].
ÊÊ

\bibitem{Arcadi:2013qia} 
  G.~Arcadi, Y.~Mambrini, M.~H.~G.~Tytgat and B.~Zaldivar,
ÊÊarXiv:1401.0221 [hep-ph].
ÊÊ
  
\bibitem{Aad:2013yqp} 
  G.~Aad {\it et al.}  [ATLAS Collaboration],
  New J.\ Phys.\  {\bf 15}, 043009 (2013)
  [arXiv:1302.4403 [hep-ex]].

\bibitem{Aad:2012qua} 
  G.~Aad {\it et al.}  [ATLAS Collaboration],
  Phys.\ Lett.\ B {\bf 719}, 299 (2013)
  [arXiv:1212.5409 [hep-ex]].

\bibitem{Ko:2013zsa} 
  P.~Ko, Y.~Omura and C.~Yu,
  JHEP {\bf 1401}, 016 (2014)
  [arXiv:1309.7156 [hep-ph]].

\bibitem{Chang:2013lfa} 
  C.-F.~Chang, E.~Ma and T.-C.~Yuan,
  arXiv:1308.6071 [hep-ph].

\bibitem{Ramos:2013wea} 
  R.~Ramos and M.~Sher,
  arXiv:1312.0013 [hep-ph].

\bibitem{Davoudiasl:2014mqa} 
  H.~Davoudiasl, W.~J.~Marciano, R.~Ramos and M.~Sher,
  arXiv:1401.2164 [hep-ph].

\bibitem{Czakon:2013goa} 
  M.~Czakon, P.~Fiedler and A.~Mitov,
  Phys.\ Rev.\ Lett.\  {\bf 110}, 252004 (2013)
  [arXiv:1303.6254 [hep-ph]].

\bibitem{ATLASluminosity}
For instance, see \url{https://twiki.cern.ch/twiki/} \\
{\tt bin/view/AtlasPublic/LuminosityPublicResults}

\bibitem{Branco:2011iw} 
  G.~C.~Branco, P.~M.~Ferreira, L.~Lavoura, M.~N.~Rebelo, M.~Sher and J.~P.~Silva,
  Phys.\ Rept.\  {\bf 516}, 1 (2012)
  [arXiv:1106.0034 [hep-ph]].

\bibitem{Peskin}
  For a convenient reference, see M.~E.~Peskin and D.~V.~Shroeder, 
  {\em An Introduction To Quantum Field Theory}, Chapter 21,
  Perseus Books Publishing (1995).

\bibitem{Chatrchyan:2013faa} 
  S.~Chatrchyan {\it et al.}  [CMS Collaboration],
  arXiv:1312.7582 [hep-ex].

\bibitem{Cheung:2009su} 
  C.~Cheung, J.~T.~Ruderman, L.~-T.~Wang and I.~Yavin,
  JHEP {\bf 1004}, 116 (2010)
  [arXiv:0909.0290 [hep-ph]].
         
\bibitem{Alloul:2013bka} 
  A.~Alloul, N.~D.~Christensen, C.~Degrande, C.~Duhr and B.~Fuks,
  arXiv:1310.1921 [hep-ph].

\bibitem{Alwall:2011uj} 
  J.~Alwall, M.~Herquet, F.~Maltoni, O.~Mattelaer and T.~Stelzer,
  JHEP {\bf 1106}, 128 (2011)
  [arXiv:1106.0522 [hep-ph]].
  
\bibitem{Sjostrand:2006za} 
  T.~Sjostrand, S.~Mrenna and P.~Z.~Skands,
  JHEP {\bf 0605}, 026 (2006)
  [hep-ph/0603175].

\bibitem{deFavereau:2013fsa} 
  J.~de Favereau, C.~Delaere, P.~Demin, A.~Giammanco, V.~Lematre, A.~Mertens and M.~Selvaggi,
  arXiv:1307.6346 [hep-ex].

\bibitem{btagging}
  CMS Collaboration [CMS Collaboration],
  ``Performance of b tagging at sqrt(s)=8 TeV in multijet, ttbar and boosted topology events,''
  CMS-PAS-BTV-13-001.

\bibitem{Chatrchyan:2012jua} 
  S.~Chatrchyan {\it et al.}  [CMS Collaboration],
  JINST {\bf 8}, P04013 (2013)
  [arXiv:1211.4462 [hep-ex]].

\bibitem{CMS:semilep}
  CMS Collaboration [CMS Collaboration],
  ``Top pair cross section in e/mu+jets at 8 TeV,''
  CMS-PAS-TOP-12-006.

\bibitem{Chatrchyan:2013xza} 
  S.~Chatrchyan {\it et al.}  [CMS Collaboration],
  arXiv:1307.4617.

\bibitem{Agashe:2012fs} 
  K.~Agashe, R.~Franceschini, D.~Kim and K.~Wardlow,
  Phys.\ Dark Univ.\  {\bf 2}, 72 (2013)
  [arXiv:1212.5230 [hep-ph]].
        
\end{thebibliography}
\end{document}